# Supergravity from Gauge Theory

**Evan Berkowitz**[*][†]
*E-mail:* berkowitz2@llnl.gov
*Nuclear and Chemical Sciences Division, Physical and Life Sciences Directorate*
*Lawrence Livermore National Laboratory*

*for MCSMC, the Monte Carlo String and M-theory Collaboration.*

Gauge/gravity duality is the conjecture that string theories have dual descriptions as gauge theories. Weakly-coupled gravity is dual to strongly-coupled gauge theories, ideal for lattice calculations. I will show precision lattice calculations that confirm large-N continuum D0-brane quantum mechanics correctly reproduces the leading-order supergravity prediction for a black hole's internal energy—the first leading-order test of the duality—and constrains stringy corrections.

---

[*]Speaker.

[†]This work was performed under the auspices of the U.S. Department of Energy by LLNL under Contract No. DE-AC52-07NA27344. This research was partially supported by the LLNL Multiprogrammatic and Institutional Computing program through a Tier 1 Grand Challenge award. This is document LLNL-PROC-699357. Associated slides are available [1]. This work is based on Refs. [2] and [3], from which the figures are taken. I thank my collaborators for their helpful feedback and do not thank United Airlines for a very unpleasant four hour delay (three hours on the tarmac) on the way home from LATTICE 2016 which prompted me to watch Independence Day and The Lego Movie, and to begin writing these proceedings.

## 1. Introduction

The construction of a theory of quantum gravity is an outstanding theoretical problem. The gauge-gravity duality conjecture is the notion that gravity is exactly equivalent to a gauge theory. If true, it would provide a path by which to define quantum gravity nonperturbatively.

The duality connects strong and weak couplings—strongly coupled gravity corresponds to weakly coupled gauge theory and vice versa—and thus, if true, provides a useful tool by which we can further our knowledge of nonperturbative regimes via perturbative calculations on the other side of the duality. Probably the most famous example is the KSS result[4], which gives a nontrivial result for $\eta/s$ in a strongly-coupled conformal theory by an easy gravity calculation.

However, the duality is currently still a conjecture, albeit well-motivated and well-supported. It is therefore useful to test the correspondence by computing related nontrivial quantities on both sides of the duality and comparing. One may find a mismatch, which would serve as a counterexample and falsify the conjecture. Performing such a test is not trivial, because it is typically beyond our ability to compute quantities on both sides of the duality, as strong coupling is a hindrance to analytic results.

However, lattice methods can give us access to nonperturbative physics on the gauge-theory side. At strong gauge coupling, the gravity side will be analytically tractable, and we can have reliable independent results that can be compared. Thus, lattice techniques provide a route for testing the duality.

The standard lore is that weakly-coupled semiclassical (super)gravity corresponds to the large-$N$ strong-coupling limit of the gauge theory. Moving away from the strong-coupling limit introduces classical stringy corrections (finite string length effects) on the gravity side, and moving away from large-$N$ introduces quantum string loop effects. We can control $N$, the rank of the gauge group, and the coupling in our lattice calculations and gain access to these corrections.

There are a variety of approaches to studying this duality. Dimensionally reducing $\mathcal{N}=1$ super-Yang–Mills (SYM) in 10D yields $(p+1)$D maximally supersymmetric YM, which should be dual to type II string theory about a black $p$-brane background [5]. Reducing to 0 spatial dimensions yields maximally-supersymmetric D0-brane quantum mechanics (D0QM)[5, 6, 7, 8], or the BFSS Matrix Model, after Banks, Fischler, Shenker, and Susskind. We study D0QM nonperturbatively with the hope of comparing with known stringy results.

In particular, type IIA superstring theory makes definite predictions for the internal energy $E$ of a black hole as a function of temperature,[5]

$$\frac{E}{N^2} = \frac{E_0(T)}{N^0} + \frac{E_1(T)}{N^2} + \mathcal{O}\left(\frac{1}{N^4}\right) \quad (1.1)$$

$$E_0(T) = a_0 T^{2.8} + a_1 T^{4.6} + a_2 T^{5.8} + \cdots \quad (1.2)$$

$$E_1(T) = b_0 T^{0.4} + b_1 T^{2.2} + \cdots \quad (1.3)$$

where $T$ is the dimensionless temperature $\lambda^{-1/3}T$ where $\lambda = g_{YM}^2 N$ is the dimensionful 't Hooft coupling. The regime we study is the 't Hooft limit—$N \to \infty$ with $\lambda^{-1/3}T$ fixed, where $E \sim N^2$—where according to the conjecture the internal energy of a bunch of D0 branes should agree with the black 0-brane mass given by (1.1). SUGRA yields an exact value for $a_0$ which numerically is 7.41. Matching to effective field theory also gives a value for $b_0$, $-5.77$ [9].

We can try to verify these values, as well as the powers of the dimensionless temperature. Our results reproduce $a_0$ with good precision, $b_0$ with large errors, and coefficients subleading in $T$ which are unknown on the gravity side. We can also try to reproduce the exponent of the subleading-in-$T$ terms—that is, 4.6 and 5.8 for $E_0$, which we can do with some success, though mild assumptions from the gravity side are needed. The extraction of the leading power of $E_0$, 2.8, from the gauge-theory side remains an outstanding challenge for the future.

## 2. D0 Brane Quantum Mechanics

D0-brane quantum mechanics is a maximally supersymmetric 0+1D gauge theory given by

$$L = \frac{1}{2g_{YM}^2} \text{Tr} \left\{ (D_t X_M)^2 + [X_M, X_M']^2 + i\bar{\psi}\gamma^{10} D_t \psi + \bar{\psi}^\alpha \gamma^M [X_M, \psi] \right\} \quad (2.1)$$

where $M$ runs from 1 to 9 and $\alpha$ from 1 to 16. The bosonic matrices $X$ and fermionic matrices $\psi$ are $N \times N$ and live in the adjoint so that $D_t \cdot = \partial_t \cdot -i[A_t, \cdot]$.

This theory has an obvious nonperturbative definition—a lattice formulation—and is quantum mechanical, so that it is unitary by construction.

A black 0-brane (referred to as a black hole when there is no risk of confusion) in this theory is a set of generic, nonperturbative matrices $X$. The physical picture is that the eigenvalues of those matrices are the coördinates of the D0 branes, while the off-diagonal elements correspond to the stringy connections between the D0s[6].

Because the potential term is proportional to $[X_M, X_M']^2$, sets of block-diagonal matrices correspond to decoupled systems. These flat directions persist quantum mechanically, so at large $N$ this theory appears to be a second quantized theory[7]—one can independently create an arbitrarty number of distinct systems and then subsequently allow them to interact. So, one can describe, for example, two black holes by partitioning $X$ into two, or one black hole and one D0 of radiation.

If this bunch of interacting D0-branes is really dual to a black hole, it must evaporate. Such a bunch has been known to be unstable for many years[10]. Recently, it has been shown that its evaporation has negative specific heat—just like a real black hole [11, 12]. However, from a more pragmatic point of view, this raises a question: can we reliably study a metastable state via Monte Carlo calculations? The answer is yes—the physical evaporation timescale and the Monte Carlo timescale for finding a instability can both be derived from phase-space arguments and can be seen to be exponentially large in $N$. Thus, we can stabilize the simulation by taking $N$ large enough unless we are exponentially unlucky. This allows us to guarantee that we land in the right phase, which is not guaranteed if one introduces a stabilizing mass deformation.

## 3. Lattice Calculation

Taking $N$ to be large is crucial for physical and numerical stability, but is also important for performing the desired test of duality—the supergravity results only strictly hold in the large-$N$ limit (though the form of the $1/N$ corrections is known) around the background which corresponds to a single bunch of eigenvalues. Finally, to compare our gauge theory calculations with SUGRA it is crucial to extrapolate our lattice results to the continuum limit, for the duality is between the two continuum theories. No previous work handled both of these systematic concerns. Additionally, previous work [13, 14, 15, 16, 17, 18, 19, 20] was unable to fit the leading coefficient $a_0$, but instead set it to its known value of 7.41 and fit the subleading behavior. Our calculation properly controls the large-$N$ and continuum extrapolations, and yields precise enough values to fit $a_0$.

The details of our discretization and ensembles are described in detail in Ref. [2]. Numerically, D0-brane quantum mechanics is nice, because it has a dimensionful coupling 't Hooft coupling $\lambda$ and no other dimensionful quantities, so scale setting is quite easy and there are no fine-tunings. Furthermore, having only 1 spacetime dimension renders the computation relatively cheap.

We run each Monte Carlo ensemble for many thousands of steps, and can perform reliable estimates of all observables. We found that very large statistics was required for a stable fit—not due to thermalization, but because there can be long-lived fluctuations that can bias the mean if the Monte Carlo sample is too small.

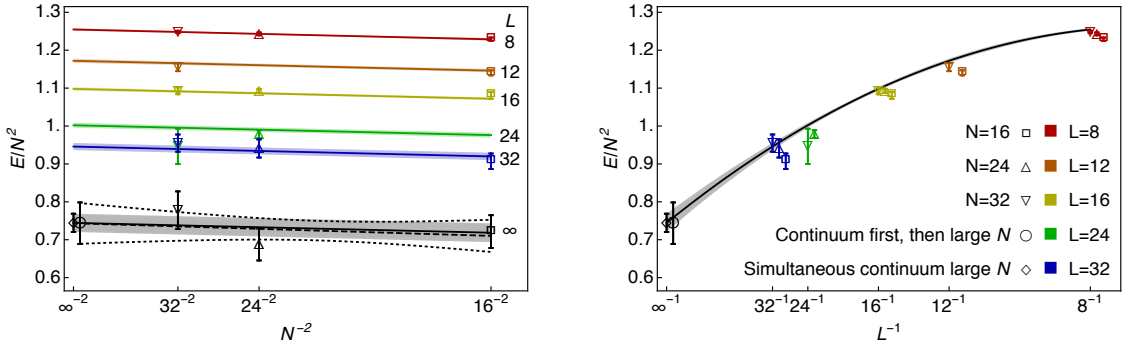

**Figure 1:** Left: A simultaneous extrapolation in $N$ and $L$ (colored lines and bands, and the black diamond) for $T = 0.5$ compared to extrapolation in $N$ (dashed line with dotted band, and the black circle) following extrapolation in $L$ (black points at finite $N$). Right: the same simultaneous surface of extrapolation, showing the quadratic dependence on $L^{-1}$. The black line and band are the large-$N$ limit at finite lattice spacings.

To reach the continuum large-$N$ result, we tried two approaches. First, we took independent continuum limits for each $N$ we studied, and subsequently took the large-$N$ extrapolation of those results. The only pitfall of this procedure is that for a successful linear extrapolation, one must make sure to be in the region of lattice spacing where $\mathcal{O}(a^2)$ effects are negligible. The linear extrapolation of coarse lattice spacings can deceptively appear reliable: going to finer spacings shows a dramatic departure and that the coarse lattices have important effects second-order in the lattice spacing. To avoid this potential issue we perform a quadratic extrapolation to the continuum. This two-step extrapolation performs acceptably, but not as well as a simultaneous extrapolation in lattice spacing and $N$.

For a given temperature we can adjust $N$ and $L$ (the number of lattice sites, which is like the inverse of the lattice spacing) independently, and fit simultaneously to

$$\frac{E}{N^2} = e_{00} + \frac{e_{01}}{L} + \frac{e_{02}}{L^2} + \frac{e_{10}}{N^2} \tag{3.1}$$

which, incorporating both the quadratic lattice spacing effects and the leading large-$N$ correction, allows us to cleanly fit all of our measured values for $E/N^2$. This procedure reliably produces the same central value with smaller error bars, as shown in Fig. 1. At a given temperature, the resulting continuum values $e_{00}$ and $e_{10}$ are the respective values of the functions $E_0(T)$ and $E_1(T)$. So, we can constrain both the leading-in-$N$ behavior and the first correction.

## 4. D0-Brane Quantum Mechanics and Supergravity Agree

With reliable fits determining the continuum values $e_{00}$ and $e_{10}$ across a range of temperatures, we can test to see if the functions $E_0(T)$ and $E_1(T)$ determined from gauge theory match what is known from the gravity side.

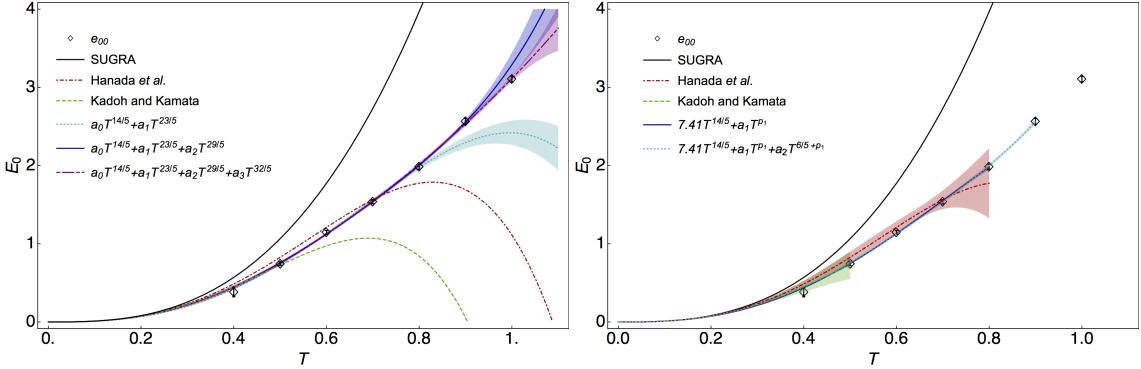

**Figure 2:** Our values of $e_{00}$ and fits for different forms of $E_0(T)$. Left: fits of coefficients only. Right: fits that fix the leading term and determine the subleading coeffients and powers. For comparison we show results from Refs. [17] and [19]. The solid black line is the SUGRA result.

In Fig. 2 we show our lattice determination of $e_{00}$ and a variety of fits. We performed two different kinds of test. First, assuming the known form of $E_0(T)$ we fit the coefficients $a_i$. We find that over the temperature range of our data care is needed to ensure there is no systematic tension distorting the coefficients—to incorporate all of our points at least three terms are needed. But, once this is handled, we get a very good, reliable fit—$a_0 = 7.4 \pm 0.5$, $a_1 = -9.7 \pm 2.2$, and $a_2 = 5.6 \pm 1.8$ with a $\chi^2/\text{DOF} = 2.6/3$. This fit is shown in blue in the left panel of Fig. 2. That we recover the correct value of $a_0$ is nontrivial—it could have differed from the SUGRA value of 7.41 and falsified the correspondence. We also find $a_1$ to be compatible with the result from [17].

In the right panel we show a test of a different sort. Instead of assuming the form of $E_0(T)$ is correct, we assume only the leading behavior and try to fit both the coefficients and powers of the subleading terms, as is done in eg. [17, 19]. Knowing that our data spans a temperature range where the third term is important means that we should not expect a fit with only one subleading term to succeed. Indeed, if we fit the form $7.41T^{2.8} + a_1 T^{p_1} + a_2 T^{p_1+1.2}$ we find $p_1 = 4.6 \pm 0.3$, $a_1 = -10.2 \pm 2.4$, and $a_2 = 6.2 \pm 2.6$ with $\chi^2/\text{DOF} = 2.6/3$. It is comforting that this determination of $a_1$ and $a_2$ is nicely compatible with our other determination. It would be preferable to not incorporate knowledge of the gravity side by assuming that the exponent of the third term $p_2$ was $p_1 + 1.2$, but without this assumption one finds a degenerate best fit of $p_1 = p_2 \neq 4.6$. A reliable independent determination would require additional data points at low temperature where $T^{4.6}$ and $T^{5.8}$ differed noticeably.

Of course, the ultimate goal is to check that the gauge theory reproduces the SUGRA result $7.41T^{2.8}$ with no assumptions from gravity. While we have shown that the gauge theory indeed produces $a_0 \approx 7.41$, this demonstration relied on knowing the power 2.8. A complete demonstration without this assumption remains an outstanding problem.

We have access to more than just the leading-in-$N$ behavior, however. In fact, we can make a comparable test at $\mathcal{O}(N^{-2})$, as the low-temperature behavior there is known as well and our calculation yields values for $E_1(T)$ via the fit parameter $e_{10}$.

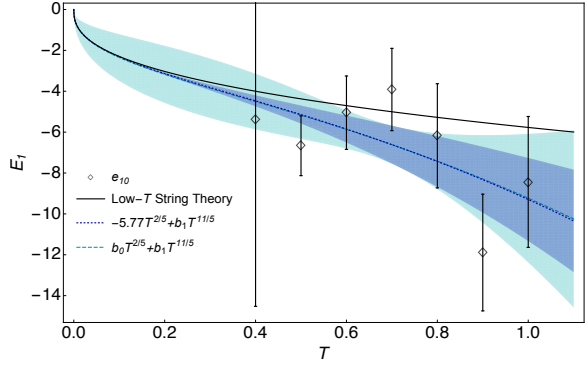

**Figure 3:** Our values for $e_{10}$ and two fits, as well as the low-temperature stringy prediction.

We show the results of trying to fit the coefficients only, assuming the polynomial form. We find $b_0 = -5.8 \pm 3.0$, compatible with the known value of $-5.77$, and loosely constrain $b_1 = -3.4 \pm 5.7$. Clearly, additional precision is needed to make a stringent comparison or to make a prediction for the analytic result on the gravity side. Nevertheless, the agreement between our value of $b_0$ and the known gravity-side result is encouraging, and jibes with previous continuum low-temperature small-$N$ results[19].

## 5. Conclusions and Outlook

By performing a strong-coupling, nonperturbative lattice calculation of D0-brane quantum mechanics, we can try to test the conjectured equivalence of gauge theory and gravity.

At leading order in $N$, the equivalence passes the test—the continuum gauge theory produces the same behavior as SUGRA. This conclusion relies on knowledge of the temperature dependence—removing the final assumption that at low temperature the black hole internal energy scales like $T^{2.8}$ remains an open challenge. Continuum knowledge of corrections in $N$ remains too imprecise to draw a strong conclusion.

Further questions remain. Can we verify emergent spacetime far from the eigenvalue bunch? Can we reach the low-temperature limit and see massless Hawking radiation in a near-thermal spectrum? Does the black hole have a firewall? Exciting horizons lie ahead.